\documentclass[english]{llncs}
\usepackage{makeidx}  

\usepackage{times}
\usepackage[]{fontenc}
\usepackage[latin1]{inputenc}
\usepackage{babel}

\usepackage{ifpdf}
\ifpdf
\usepackage[pdftex]{graphicx}
\usepackage{epstopdf}
\else
\usepackage{graphicx}
\fi

\usepackage{cite}

\usepackage{subfigure} 

\usepackage{amsmath}   
\usepackage{amssymb}
\usepackage{array}

\begin{document}

\newcommand{\DDDDDD}{6D}
\newcommand{\DDDD}{o2D}
\newcommand{\DDD}{3D}
\newcommand{\DD}{2D}
\newcommand{\R}{\ensuremath{\mathbb{R}}}
\newcommand{\C}{\ensuremath{\mathbb{C}}}

\frontmatter          
\pagestyle{empty}  
\mainmatter              

%
\title{Towards \DDD~Ultrasound Image Based Soft Tissue Tracking: a Transrectal Ultrasound Prostate Image Alignment System.}

\author{Michael~Baumann\inst{1,2}, Pierre~Mozer\inst{1,3}, Vincent~Daanen\inst{2}, and~Jocelyne~Troccaz\inst{1}}
\institute{Université J. Fourier, Laboratoire TIMC, Grenoble, France; CNRS, UMR 5525; Institut National Polytechnique de Grenoble.\and Koelis SAS, 5 av. du Grand Sablon, 38700 La Tronche, France.\and La Pitié-Salpêtrière Hospital, Urology Department, 75651 Paris Cedex 13, France.\\ \email{michael.baumann@imag.fr}}
\authorrunning{M. Baumann et al.}   

\maketitle

\begin{abstract}
The emergence of real-time \DDD~ultrasound (US) makes it possible to consider image-based tracking of subcutaneous soft tissue targets for computer guided diagnosis and therapy. We propose a \DDD~transrectal US based tracking system for precise prostate biopsy sample localisation. The aim is to improve sample distribution, to enable targeting of unsampled regions for repeated biopsies, and to make post-interventional quality controls possible. Since the patient is not immobilized, since the prostate is mobile and due to the fact that probe movements are only constrained by the rectum during biopsy acquisition, the tracking system must be able to estimate rigid transformations that are beyond the capture range of common image similarity measures. We propose a fast and robust multi-resolution attribute-vector registration approach that combines global and local optimization methods to solve this problem. Global optimization is performed on a probe movement model that reduces the dimensionality of the search space and thus renders optimization efficient. The method was tested on 237 prostate volumes acquired from 14 different patients for \DDD~to \DDD~and \DDD~to orthogonal \DD~ slices registration. The \DDD-\DDD~version of the algorithm converged correctly in 96.7\% of all cases in 6.5s with an accuracy of 1.41mm (r.m.s.) and 3.84mm (max). The \DDD~to slices method yielded a success rate of 88.9\% in 2.3s with an accuracy of 1.37mm (r.m.s.) and 4.3mm (max).
\end{abstract}

\section{Introduction}
Computer-guidance for medical interventions on subcutaneous soft tissue targets is a challenging subject, since the target tracking problem is still not satisfactorily solved. The main difficulties are caused by the elasticity, mobility and inaccessibility of soft tissues. With \DDD~US a real-time volume imaging technology became available that provides enough spatial tissue information to make image-based tracking possible. Image-based tracking is essentially a mono-modal image registration problem with a real-time constraint. The primary task is to find the physical transformation $T$ in a transformation space $\mathcal{T}$ between two images of the same object. The choice of $\mathcal{T}$ depends on the underlying physical transformation (e.g. rigid, affine or elastic) and the requirements of the target application. An extensive review on registration methods is given in \cite{Zitova03Survey}.\\ \indent
Nowadays, research on mono-modal \DDD~US registration of soft tissue images focusses on rapid deformation estimation. Most studies in this domain, however make the implicit assumption that the rigid part of the transformation to estimate is either small or known. Confronted with combinations of large rigid transformations and elastic deformations, the proposed solutions fail without rigid pre-registration. For many clinical applications large rigid transformations can be avoided by immobilizing both the patient and the US probe. In the case of interventions without total anesthesia this however causes considerable patient discomfort. Moreover, it is sometimes impossible to fix the US probe, e.g. when the probe serves as a guide for surgical instruments. The respiratory and the cardiac cycle can be additional sources of tissue displacements. In all these cases it is necessary to identify the rigid part of the transformation before carrying out image-based deformation estimation.\\ \indent
Estimation of large rigid transformations is basically a global optimization problem since common similarity measures exhibit search-friendly characteristics (e.g. convexity) only in a small region near the global solution. The computational burden of global optimization in a 6-D rigid transformation space is prohibitive for tracking tasks. \cite{Gueziec97hashing,Eadie04indexblock} propose to reduce the intra-interventional computation time of global searches by pre-computing a feature-based index hash table. During intervention, similarity evaluation is replaced by computation of the geometric index followed by a fast data-base look-up. In the context of US image tracking, this approach has the disadvantage of relying on feature extraction, which often lacks robustness when confronted with partial target images, speckle and US shadows. Also, it cannot reduce the complexity of the optimization problem and pre-computation time is not negligible.\\ \indent
Relatively few investigations involving \DDD~US image based tracking of soft tissues have been reported. In the context of respiratory gated radiation treatment, \cite{Sawada04Radiation} acquire a localized \DDD~US reference image of the liver or the pancreas in breath-hold state and register it rigidly with the treatment planning CT volume. During therapy, localized US slices of the organ are continuously compared with the reference volume using image correlation to retrieve the planning position of the organ. In \cite{Huang05Dynamic} real-time \DDD~US images of the beating heart are registered multimodally with a set of 4-D MR images covering the entire cardiac cycle. A localizer is used to initialize the spatial registration process while the ECG signal serves for temporal alignment. The authors achieve precise rigid registration in an overall computation time of 1 second with a mutual information based rigid registration algorithm. In both studies relative rigid movements between probe and target organ are limited to movements caused by the respiratory or cardiac cycles, which are predictable and repeatable to a certain extent.\\ \indent
The target application of this work is \DDD~transrectal ultrasound (TRUS) prostate biopsy trajectory tracking. Today, prostate biopsies are carried out using \DD~TRUS probes equipped with a guide for spring needle guns. With the current standard biopsy protocol, consisting typically of 12 regularly distributed samples, it is impossible to know the exact biopsy locations after acquisition, which makes precise biopsy-based tumor localization, quality control and targeted repeated biopsies impossible. A TRUS-based prostate tracking system would make it possible to project all sample locations into a reference image of the prostate and thus to identify the exact sampling locations.\\ \indent
Image-based prostate biopsy tracking is, however, challenging: (i) the gland moves and gets deformed under the pressure of the TRUS probe. (ii) The patient is neither immobilized nor under total anesthesia. Most patients move significantly during the biopsy procedure. (iii) Since the probe serves also to guide the rigidly attached needle, probe movements are important. Rotations around the principal probe axis of more than 180° and tilting of up to 40° are frequent. Also, the probe head wanders over the gland surface during needle placement, which leads to relative displacements of up to 3cm. The global search problem thus fully applies to prostate alignment: tracking a reference on a calibrated TRUS probe cannot solve the problem due to (i) and (ii), and it is not very success promising to minimize similarity measures on biopsy images using only fast down-hill optimizers because of (iii). In this study we propose a solution to the global search problem for TRUS prostate image tracking, which consists in a search space reduction using a probe movement model. We further identify an efficient intensity-based similarity measure for TRUS prostate images and describe a fast multi-resolution optimization framework. Finally, the robustness, accuracy, precision and performance of the method are evaluated on 237 prostate volumes from 14 patients.
\section{Methods}
\subsection{A framework for US image-based tracking}
The purpose of a tracking system is to provide the transformation between an object in reference space and the same object in tracking space at a given moment. In the case of image-based tracking, the reference space is determined by the choice of a reference image to which all subsequently acquired images will be registered. In the case of \DDD~TRUS prostate biospies, it is convenient to acquire a \DDD~US volume as reference just some minutes before the intervention.

Unfortunately, most currently available \DDD~US systems do not provide real-time access to volume data. They can, however, visualize two or three orthogonal \DD~(\DDDD) slices inside the field of view of the probe in real-time. These slices can be captured using a frame-grabber and used for registration with a previously acquired reference volume \cite{Sawada04Radiation,Huang05Dynamic}. Note that compared to \DD~US images, \DDDD~planes deliver considerably more spatial information, which potentially makes \DDD~to \DDDD~registration more robust than \DDD~to \DD~registration. In this work we will evaluate both \DDD~to \DDD~and \DDD~to \DDDD~registration for image-based tracking.

Registration algorithms can be separated into two main classes: intensity-based and feature-based algorithms. As it is challenging to define robust \textit{and} fast feature extraction algorithms for US images of the prostate, due to the low SNR of US images and the absence of clearly identifiable geometric features in the prostate, this study focuses on intensity-based approaches. Intensity-based measures are known for their robustness in presence of noise and partial image overlaps \cite{Zitova03Survey}. 

Image registration can be modeled as a minimization process of an image similarity measure that depends on a transformation $T$. There exist robust and fast algorithms for local minimization of image similarity measures. The condition for convergency to the target transformation $\hat{T}$ is that the optimizer starts from a point inside the capture range of $\hat{T}$ \cite{shekhar02ultrasoundreg}. However, the capture range of common intensity measures (e.g. the Pearson correlation coefficient (CC) or normalized mutual information (NMI)) is relatively small compared to the transformation space that can be observed for TRUS prostate biopsies. This problem can be attacked from two sides: the first approach is to extend the capture range by improving the similarity measure, while the second method consists in finding a point inside the capture range using a priori knowledge on the probe position.

Several parts of the registration approach require information about the prostate location in the reference image. For our purpose it is sufficient to set an axis-aligned bounding box on the prostate boundaries in the reference image. The bounding box has to be defined by the clinician. No bounding box is needed for the tracking images.

%
%
\subsection{Extending the Capture Range}

\textbf{Similarity Measure:} We chose CC as similarity measure since it yields a larger capture range than NMI for mono-modal US registration. Compared to sums of squared distances (SSD), it is insensitive to linear intensity transformations and is capable of detecting inverse correlations. Intensity shifts can occur due to probe pressure variation, while inverse correlations can be observed when evaluating transformations far from the physical solution, in particular for gradient magnitude images.

\textbf{Multi-resolution pyramid:} Optimizing on coarse resolution levels of a gaussian pyramid yields some important advantages: coarse levels are statistical aggregates of the original image which are free of high-frequency noise, in particular speckle noise. Once the optimization on the coarsest level is terminated, the solution will be refined on denser levels, but from a considerably better starting point. This approach not only improves the characteristics of the similarity measure by reducing noise, but also considerably speeds up registration time, as most of the optimization can be performed on low-resolution images.

\textbf{Attribute-vector approach:} The capture range can be extended by combining measures of different aspects of the images to be compared \cite{Shen02hammer,foroughi05elastic}. Since there is a strong probability that the similarity measure produces for every aspect a significant minimum near the correct solution, it is possible to amplify and widen the capture range of the solution by combining the measures. Also, it is less likely that noise-related local minima are produced at identical locations, which makes it possible to flatten them out in a combined measure. For this study we chose to evaluate the image intensity and its gradient magnitude ($I$ and $J$ are the images to be compared):
\begin{equation}
\label{equ:combinedenergy}
\begin{split}
En_{IJ}(T) := & ~(1 - CC(I, J \circ T)) \cdot (1 - CC(||\nabla I||, ||\nabla J \circ T||))
\end{split}
\end{equation}
To improve performance and since gradient intensities are highly random on noisy high-resolution images, attribute vectors are only used on low resolution levels of the image pyramid.

\textbf{Panorama images:} The pyramid-like form of the US beam and the fact that the probe also serves to guide the biopsy needle makes it unavoidable that the gland is often only partially imaged. Hence at least the reference image should contain the entire prostate; otherwise the similarity measure may yield random results when the image overlap gets too small during registration. We therefore acquire three partial prostate volumes using the following protocol: the operator first acquires one image where the prostate is centered in the US beam, and then takes two additional images with rotations of 60° around the principal axis of the probe. Care is taken to avoid deformation and US shadows. The panorama image resulting from compounding these acquisitions finally serves as reference.
%
%
%
\subsection{Finding a point in the capture range}
\subsubsection{Mechanical probe movement model:}
To estimate large transformations between images, it is necessary to find a point inside the capture range of the similarity measure. Regular sampling of a 6-D rigid transformation space using a very sparse grid size of 10 already requires $10^6$ function evaluations, which results in an unacceptable computational burden. The physical constraints exerted by the rectum on probe movements, and the fact that the probe head always remains in contact with the thin rectal wall at the prostate location lead to the following assumptions: 1) the probe head is always in contact with the prostate membrane, 2) the most important rotations occur around the principal axis of the probe, and 3) all other rotations have a rotation point that can be approximated by a unique fixed point $FP_{rect}$ in the rectum.

\begin{figure}
	\centering
		\subfigure[]{\includegraphics[width=.3\textwidth]{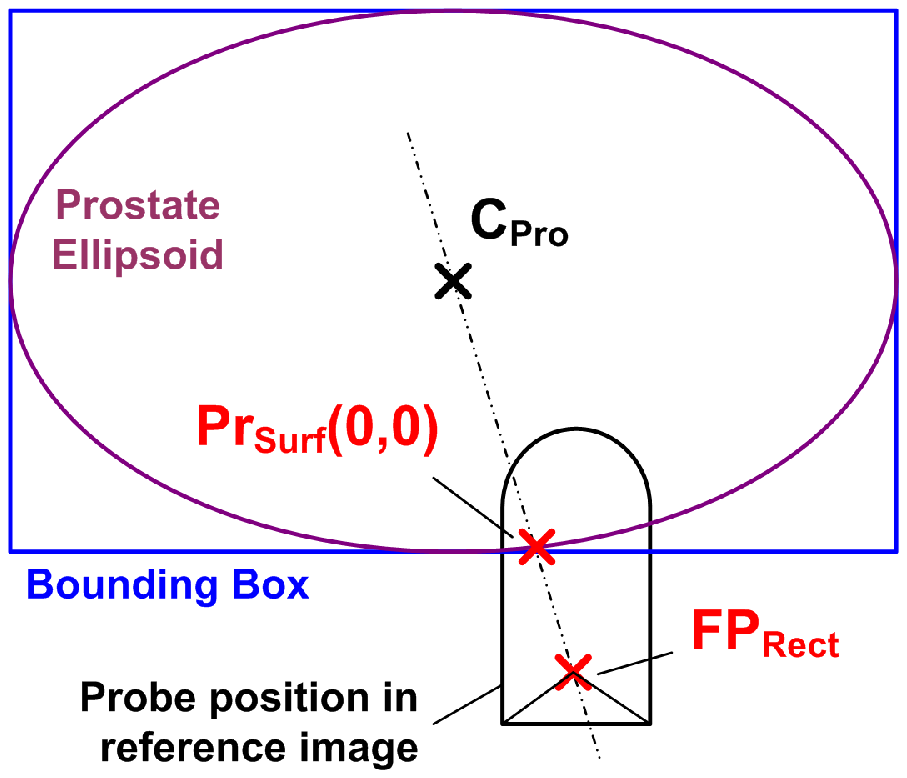}\label{fig:model1}}
		\subfigure[]{\includegraphics[width=.3\textwidth]{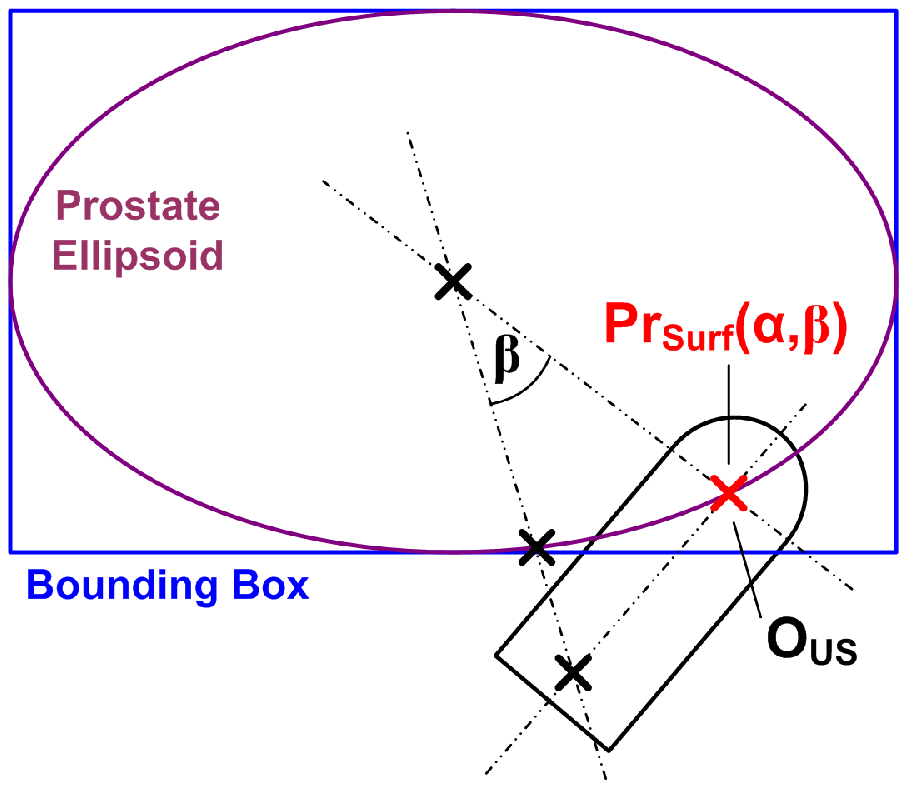}\label{fig:model2}}
		\subfigure[]{\includegraphics[width=.3\textwidth]{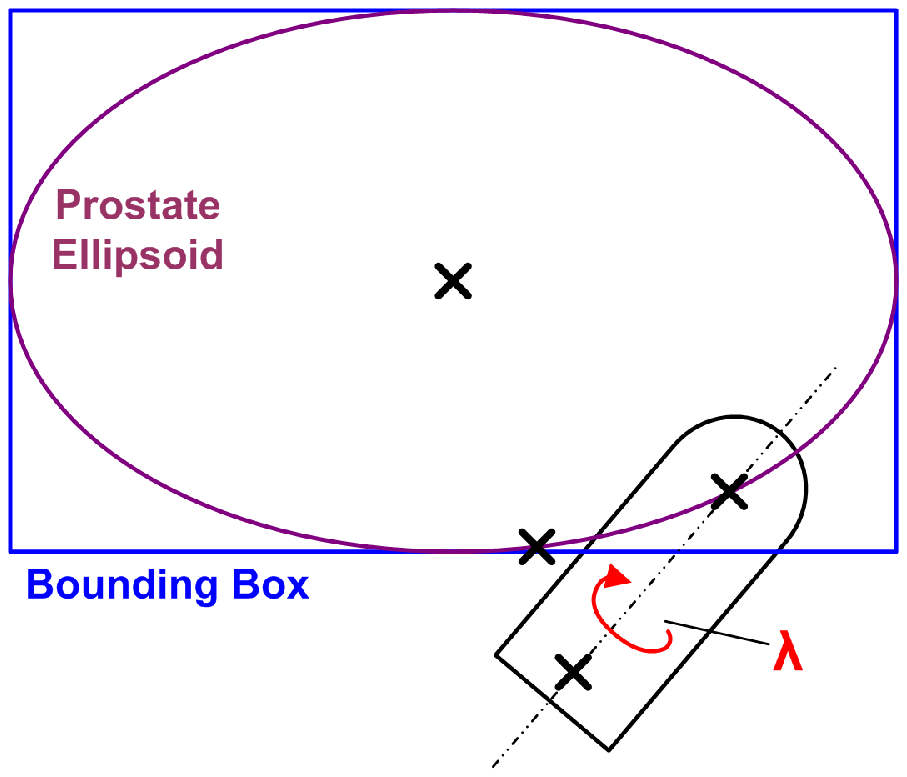}\label{fig:model3}}
		\caption{\textbf{Mechanical probe movement model in \DD:} (a) shows the computation of the search model surface origin $PR_{Surf}(0,0)$ from the prostate center $C_{Pro}$ and the (hypothetical) rectal probe fixed point $FP_{Rect}$. In (b), a \DD~polar parameterization is used to determine a surface point $PR_{Surf}(\alpha, \beta)$. The probe is then rotated and translated such that its US origin $O_{US}$ coincides with $PR_{Surf}(\alpha, \beta)$. In (c), the probe is rotated around its principal axis by an angle $\lambda$.}
\end{figure}

With these assumptions it is possible to define a probe movement model based on a prostate surface approximation, the probe position in the US image (which is known) and a rotational fixed point in the rectum. As shown in Fig. \ref{fig:model1}, the prostate surface is approximated by a bounding-box aligned ellipsoid. The ellipsoid is modeled using a \DD~polar parameterization $PR_{Surf(\alpha,\beta)}$. The origin $PR_{Surf(0, 0)}$ of the parameterization corresponds to the intersection of the line from the prostate center $C_{Pro}$ to $FP_{Rect}$. As illustrated in Fig. \ref{fig:model2}, $PR_{Surf(\alpha,\beta)}$ implements assumption 1) by determining plausible US transducer positions on the prostate surface. Assumption 3) is satisfied by requiring that the principal probe axis must always pass through $FP_{Rect}$. Finally, a rotation about the principal probe axis implements assumption 2) and thus adds a third DOF (See Fig. \ref{fig:model3}).
\subsubsection{Systematic Exploration}
The \DDD~subspace defined by the probe movement model is systematically explored using equidistant steps. To minimize the computational burden, systematic exploration is performed on the coarsest resolution level. Since the exploration grid points do not change during an intervention, it is possible to precompute and to store all resclices of the panoramic image necessary for the evaluation of the intensity measure. The rotational space around the principal axis of the probe is unconstrained (360°), while tilting ranges are limited to the maximum value determined on test data, plus a security margin. The number of steps per dimension are also experimentally determined. The five best results of the systematic exploration are stored with the constraint that all transformations respect a minimum distance between each other. If two results are too close, only the best one is stored. Next, a local search using the Powell-Brent algorithm is performed only on the coarsest pyramid level for each of the five results. The best result of the five local searches is finally used as the start point for a multi-level local optimization. The last level of the final search can be chosen in function of the desired precision and computation time. Note that compared to a single multi-level local search, five local optimizations on the coarsest level are negligible in terms of computation time. 

%
%

\section{Experiments and Results}

The presented method was validated on 237 \DDD~images of the prostate acquired during biopsy of 14 different patients. The imaging device was a GE \DDD~US Voluson 730 equipped with a volume-swept transrectal probe (GE RIC5-9). All images, except the images used for panorama image creation, were acquired immediately after a biopsy shot. Both \DDD~to \DDD~and \DDD~to \DDDD~registration were evaluated. All registrations were carried out in a post-processing step. The \DDDD~images used in the tests were not frame-grabbed but reconstructed from \DDD~images. The image resolution was $200^{3}$. The voxel side lengths varied from 0.33mm to 0.47mm. A five-level resolution pyramid was used for \DDD~to \DDD~registration; for \DDD~to \DDDD~only four levels were used. The final multi-level search was carried out from the coarsest to the third-finest level for \DDD~to \DDD, and to the second-finest level for \DDD~to \DDDD~registration. A total of 12960 grid points on the movement model were explored during a search run. Registration was carried out on a Pentium 4 with 3GHz.

\begin{figure}
	\label{fig:examples}
	\centering
		\subfigure[]{\includegraphics[width=.23\textwidth]{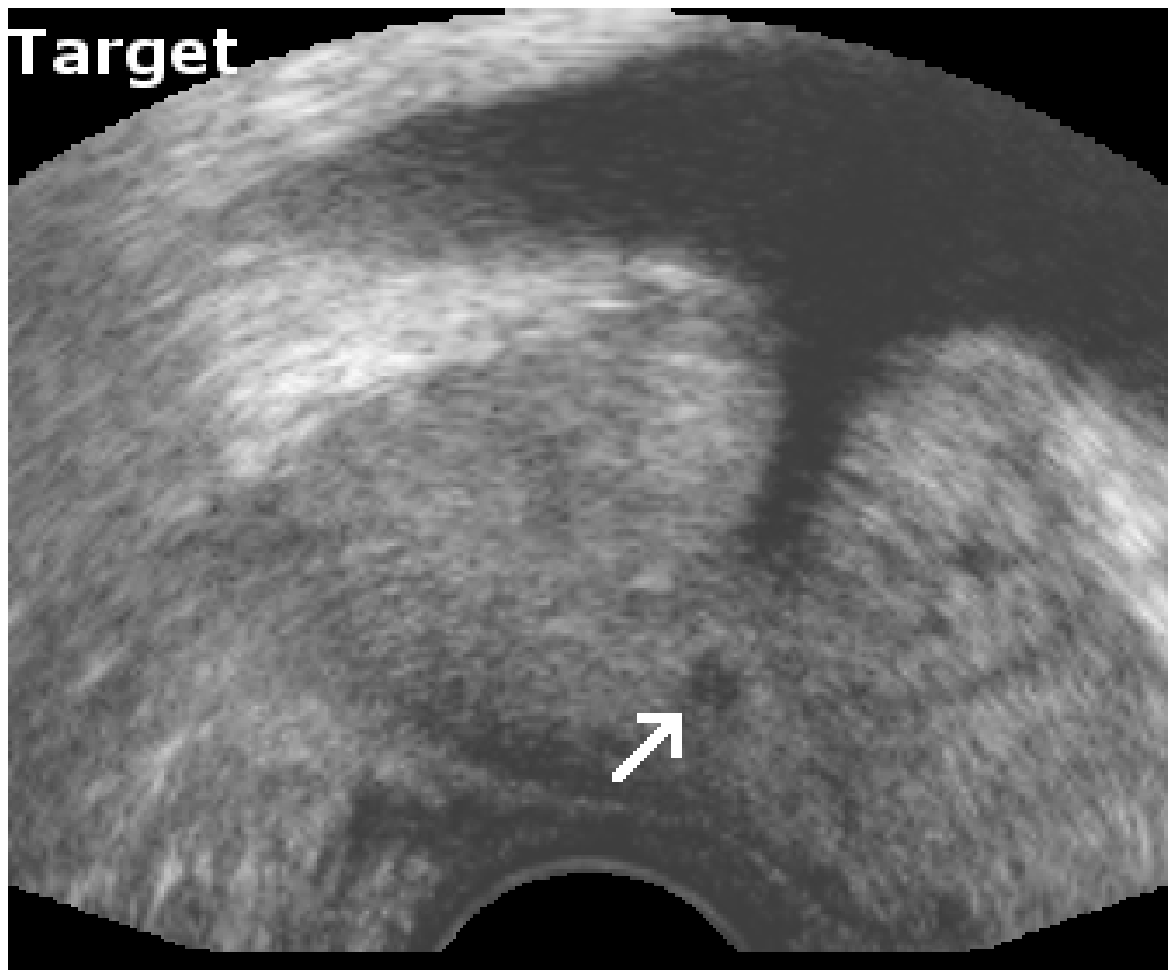}\label{fig:example1}}
		\subfigure[]{\includegraphics[width=.23\textwidth]{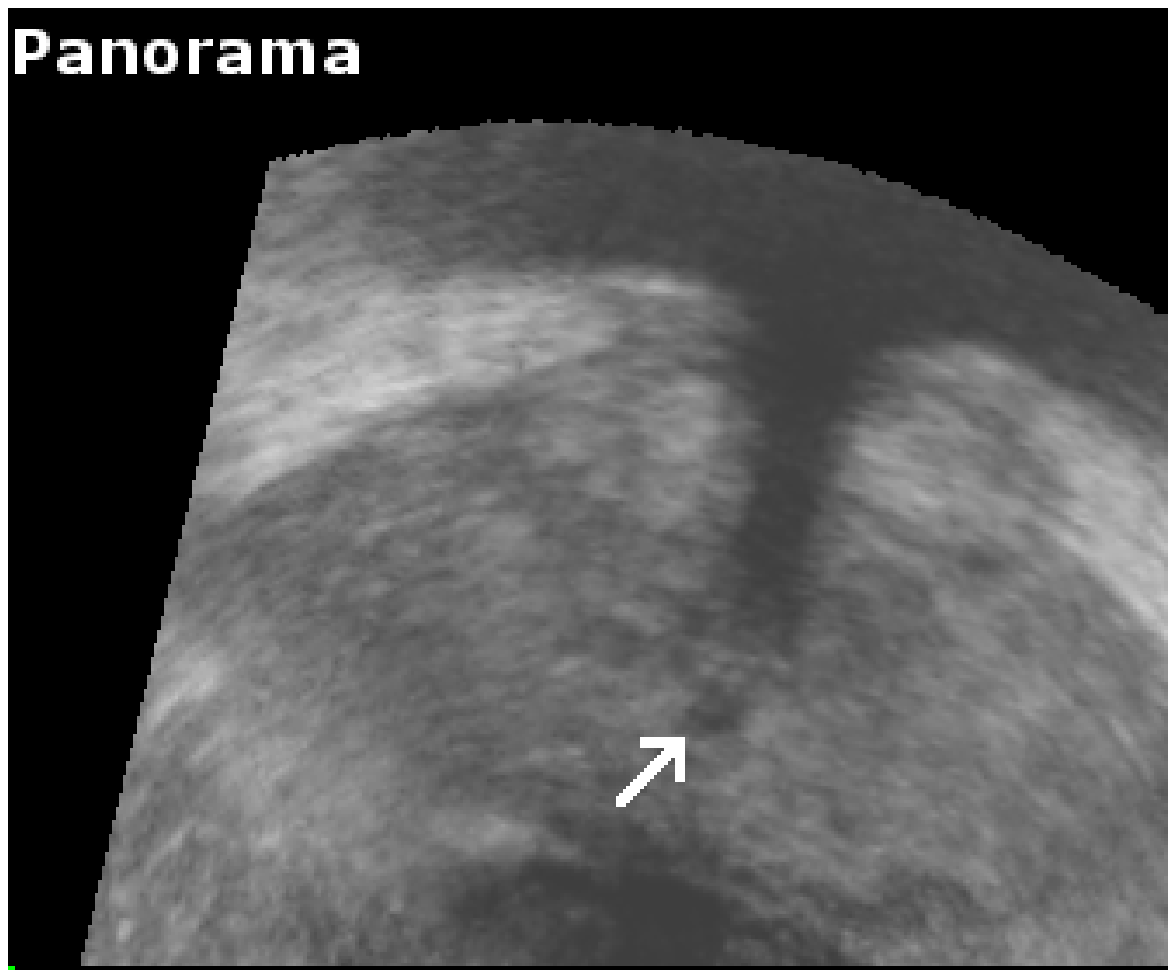}\label{fig:example2}}
		\subfigure[]{\includegraphics[width=.23\textwidth]{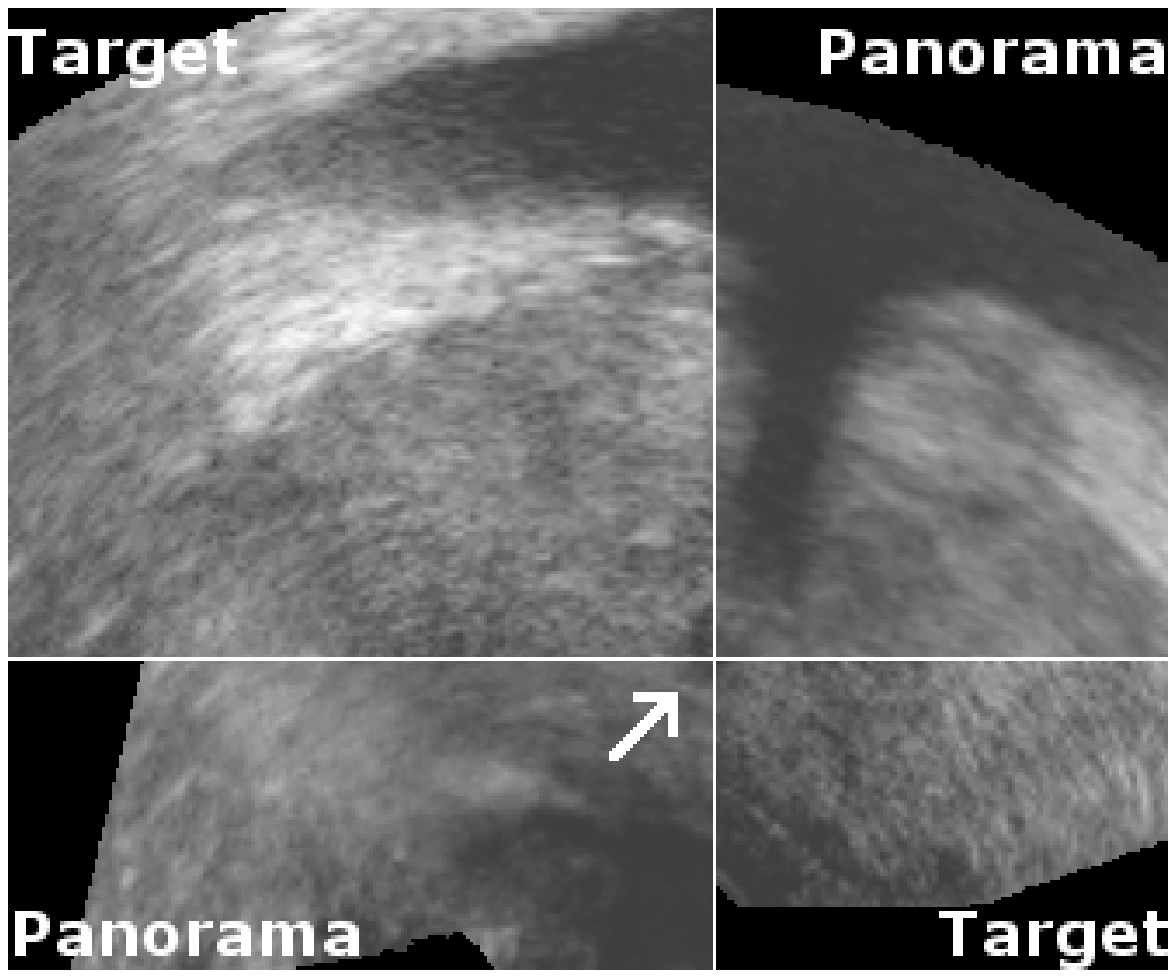}\label{fig:example3}}
		\subfigure[]{\includegraphics[width=.23\textwidth]{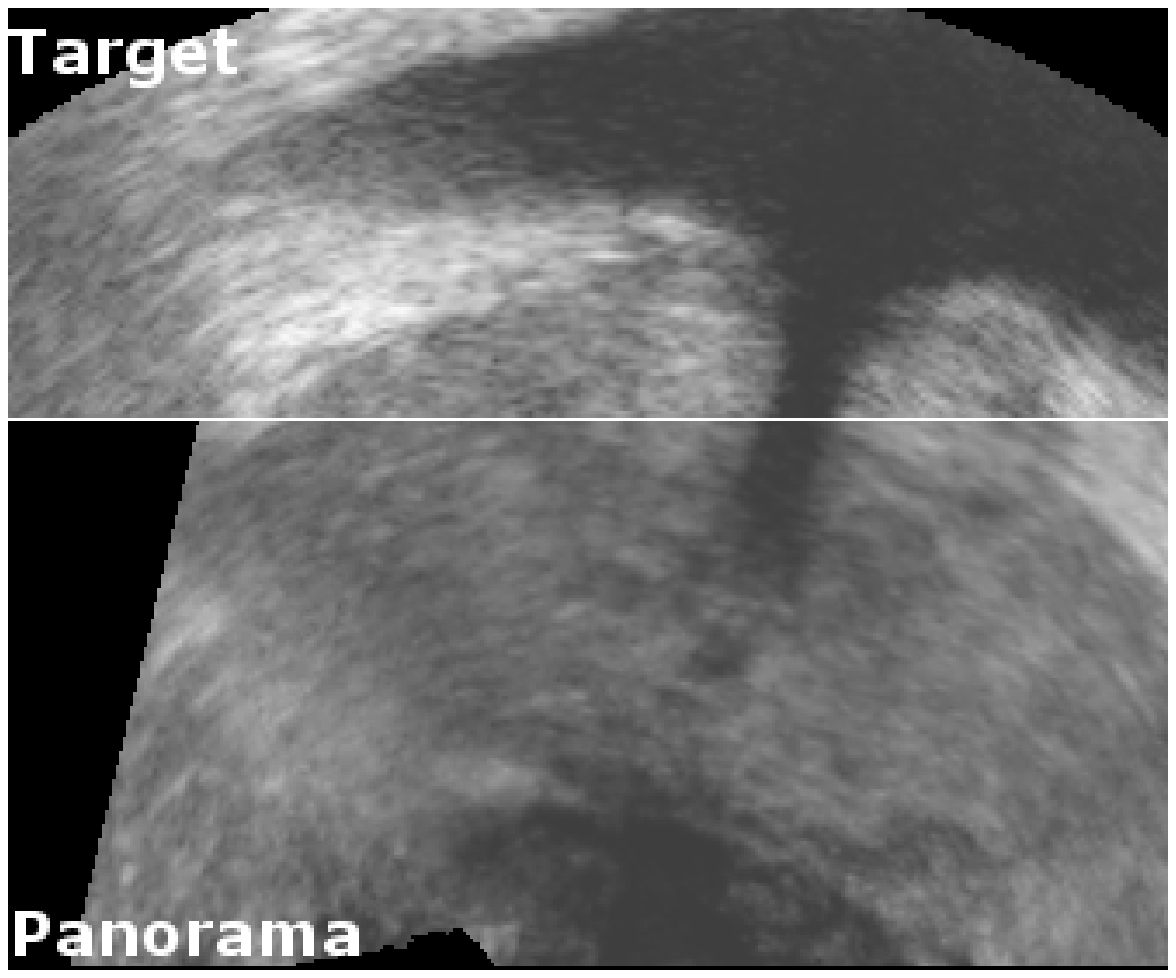}\label{fig:example4}}
		\caption{\textbf{Registration accuracy:} (a) shows the target image, and (b) the aligned panorama image. In (c) both volumes are superimposed to illustrate registration accuracy for the urethra (arrow), and (d) illustrates the registration accuracy in the upper gland.}
\end{figure}

To measure reproducibility and registration success, 10 registrations were carried out for each volume pair from slightly perturbated start points by adding noise of 2mm and 2°. This yielded 10 transformations $T_i$ that approximate the unknown rigid transformation between the prostate in both volumes. The average transformation $\overline{T}$ of the $T_i$ was computed with the method presented in \cite{gramkow01rotavg}. The euclidean distance error $\epsilon_{E}^{i}=||T_i\cdot C - \overline{T}\cdot C||$, with $C$ being the image center, and the angular error $\epsilon_{A}^{i}$, which corresponds to the rotation angle of $T_i^{-1}\cdot\overline{T}$, were used to compute the root mean square (r.m.s.) errors $\epsilon_E$ and $\epsilon_A$. A registration was considered successful if $\epsilon_{E}<2.0$mm and $\epsilon_{A}<5$ degrees, and if the result $\overline{T}$ was visually satisfactory when superimposing both volumes in a composite image (See Fig. \ref{fig:example3}).

Reconstruction accuracy evaluation was more difficult to implement since there is no straight-forward gold standard. In some images, the needle trajectories from previous biopsies were still visible. In these cases, the trajectories were manually segmented, and the angular error between corresponding needle trajectories were used to evaluate rotational accuracy. Also, some patients had significant and clearly visible calcifications inside the prostate. The distances between segmented calcifications were used to determine the translational accuracy. Tab. 1 and Fig. 2 show the results of the evaluations. 
\vspace{-10pt}
\begin{table}
\label{tab:evaluation}
\centering
\begin{tabular}{lcc}
  & \textbf{~~~~~~~~~\DDD-\DDD~~~~~~~~~} & \textbf{~~~~~~~~~\DDD-\DDDD~~~~~~~~~} \\ \hline \vspace{-5pt} \\ 
\textbf{Registration success} &  96.7\% (237)& 87.7\% (237)\\
\textbf{Average computation time} & 6.5s (237)& 2.3s (237)\\
\textbf{Angular precision $\epsilon_A$ (reproducibility, r.m.s.)} & 1.75° (229)& 1.71° (208)\\
\textbf{Euclidean precision $\epsilon_E$ (reproducibility, r.m.s.)} & 0.62mm (229)& 0.47mm (208)\\
\textbf{Needle trajectory reconstruction (r.m.s.)} & 4.72° (10)& 4.74° (9)\\
\textbf{Needle trajectory reconstruction (max)} & 10.04° (10)& 10.5° (9)\\
\textbf{Calcification reconstruction (r.m.s.)} & 1.41mm (189)& 1.37mm (181) \\
\textbf{Calcification reconstruction (max)} & 3.84mm (189)& 4.30mm (181)\\ \hline\vspace{-5pt}\\
\end{tabular}
\caption{\textbf{Test results}: Numbers in brackets indicate the number of evaluated registrations.}
\end{table}
\vspace{-20pt}

The overhead introduced by the systematic model-based exploration accounts for about 25\% of \DDD-\DDD~, and for 35\% of \DDD-\DDDD~registration time. The five optimizations on the coarsest level account for about 10\% in \DDD-\DDD, and for 20\% in \DDD-\DDDD. Panorama image pre-processing and pre-computation of the images for systematic exploration are performed before the intervention and require about one minute of computation time.
\vspace{-20pt}
\section{Discussion}
This study presents a fast and robust rigid registration framework for TRUS prostate images in the context of unconstrained patient movements, of only anatomy-constrained probe movements and of probe-induced prostate displacements. The algorithm yields reproducible results and acceptable accuracy for both \DDD-\DDD~and \DDD-\DDDD~registration. 

The success-rate of \DDD-\DDD~registration is very satisfactory, since all failures were either due to significant US shadows caused by only partial contact of the probe head with the rectal wall or by air bubbles in the US contact gel, or to an insufficient US depth with the result that parts of the gland membrane are not visible in the images. In these cases the similarity measure fails because of missing information in the image, and an algorithmic remedy probably does not exist. Additional failures can be observed for \DDD-\DDDD~registration, in particular for very small prostates, for which the coronal plane does not contain any prostatic tissue. \DDD-\DDDD~registration is also more sensible to poor image quality (e.g. low contrast), to large deformations and to partial prostate images (for which often only one plane contains prostatic tissue). Note that the presented algorithm is not very sensible to bounding box placement precision.

Computation time of local searches could be accelerated using the GPU for image reslicing (which corresponds to approximatively 95\% of the computational burden of a similarity measure evaluation), while further optimization of the systematic exploration would require parallelization of the evaluations.

The presented algorithm in particular accurately registers the prostate membranes that are distant to the probe head, and the urethra. The relatively high angular r.m.s. error observed in the needle reconstruction study can be explained with probe-related local deformations that are particularly strong at the needle entry point. We are currently working on a biomechanical gland deformation model that allows for estimation of deformations to improve the accuracy of tissue registration near the probe head.
\\
\\
\begin{scriptsize}
\textbf{Acknowledgements:} This work was supported by grants from the Agence Nationale de la Recherche (TecSan program, SMI project), from the French Ministry of Industry (ANRT agency), from the French Ministry of Health (PHRC program, Prostate-echo project) and from Koelis S.A.S., France. The clinical data were acquired at the urology department of the Pitié la Salpétrière hospital, Paris.
\end{scriptsize}
\bibliographystyle{splncs}
\bibliography{ProstateRigid}
\end{document}